\documentclass[fleqn,10pt]{wlscirep}
\title{Localisation of weakly interacting bosons in two dimensions: disorder vs lattice geometry effects}

\author[1]{Luis A. Gonz\'alez-Garc\'ia}
\author[1,2,3]{Santiago F. Caballero-Ben\'{\i}tez}
\author[1,*]{Rosario Paredes}
\affil[1]{Instituto de F\'{\i}sica, Universidad
Nacional Aut\'onoma de M\'exico, Apartado Postal 20-364, M\'exico D.
F. 01000, M\'exico}
\affil[2]{CONACYT-Instituto Nacional de Astrof\'{\i}sica, \'Optica y Electr\'onica, Calle Luis Enrique Erro No. 1, Sta. Mar\'{\i}a Tonantzintla, Pue. CP 72840, M\'exico}
\affil[3]{University of Oxford, Department of Physics, Clarendon Laboratory, Parks Road, Oxford OX1 3PU, United Kingdom}

\affil[*]{rosario@fisica.unam.mx}

\begin{abstract}

We investigate the effects of disorder and lattice geometry against localisation phenomena in a weakly interacting ultracold bosonic gas confined in a 2D optical lattice. The behaviour of the quantum fluid is studied at the mean-field level performing computational experiments, as a function of disorder strength for lattices of sizes similar to current experiments. Quantification of localisation, away from the Bose glass phase, was obtained directly from the stationary density profiles through a robust statistical analysis of the condensate component, as a function of the disorder amplitude. Our results show a smooth transition, or crossover,  to localisation induced by disorder in square and triangular lattices. In contrast, associated to its larger tunneling amplitude, honeycomb lattices show absence of localisation for the same range of disorder strengths and same lattice amplitude, while also exhibiting partial localisation for large disorder amplitudes. We also conclude that the coordination number $z$ have a partial influence on how fast this smooth transition occurs as the system size increases. Signatures of disorder are also found in the ground state energy spectrum, where a continuous distribution emerges instead of a distribution of sharp peaks proper to the system in the absence of disorder.  
 
\end{abstract}
\begin{document}

\flushbottom
\maketitle

\thispagestyle{empty}

\section*{Introduction}

Absence of transport in solids is intimately related with either, interparticle interactions or structural disorder. Here, we focus our attention in localization phenomenon induced by disorder in lattices in two dimensions. The typical model for a real electronic material that predicts localisation for finite disorder strength (Anderson transition\cite{Anderson}), states that at zero temperature ($T=0$) this occurs for dimensions $d\leq2$\cite{Abrahams}. In other words, a metallic system has always localised states for any disorder strength in $d\leq2$. Several refinements to those predictions have been formulated for isotropic media \cite{Thouless,Licciardello,Lee1,Gornyi,Volthardt,WolfeRev} having a random distribution of scatterers. However, since Anderson localization is indeed a universal phenomenon, it than can be found also in anisotropic media, and in different situations than those belonging to the context of electronic transport. For instance, propagation of light in solids\cite{Pine,Kao} in either anisotropic or orientationally ordered scatterers have been explored. In the condensed matter scenario, it is difficult to isolate the effects of the underlaying lattice structure against those associated to interparticle interactions. In contrast, ultracold neutral gases are perhaps the most versatile and simple setting to explore new possibilities in the study of transport properties, as compared to the analogous real materials with electrically charged particles\cite{UCA-disorder,Sanchez-Palencia0,Beeler,Allard}. Neutral ultracold gases confined in an optical lattice allow to have each aspect related to either dynamical or stationary properties, to be externally controlled with high precision \cite{Lewenstein}. Mott insulating transitions addressed theoretically in boson \cite{Jaksch1,Jaksch2,Fisher}, and fermion \cite{Jaksch2,Liu, Goldman} cases, have been also experimentally tested in both, Fermi~\cite{Jordens} and Bose \cite{Porto} samples. Even more, Bose glass phases \cite{Deissler, Meldgin} and Anderson localization and have been realized in bosonic \cite{Roati, Sanchez-Palencia, White, Plodzien, Piraud, Jendrzejewski} and fermionic systems\cite{Kondov, DeMarco}. More recently, the interplay of disorder and interactions has been tested in Bose~\cite{Carleo,Bloch} and Fermi~\cite{Kondov2} gases confined optical lattices. In particular, in that experiment, the difficulty to reach equilibrium configurations caused by disorder and the interplay with strong two body on-site interactions were explored. From the theoretical perspective, there has been an extensive amount of work addressing the study of transport in disordered  optical lattices~\cite{Lewenstein, LSanchez-Palencia, Scarola, Vettchinkina, Kartsev}. Additionally,  proposals to investigate fractal structure induced by disorder~\cite{Fractal} and measurement of system replicas simultaneously~\cite{Replicas} have been put forward. The interplay of disorder with spin-orbit coupling~\cite{SODisorder}, spin degrees of freedom~\cite{SpinorDisorder,SpinorDisorder2}, supersolidity~\cite{SSDisorder}, and Dirac fermions~\cite{DiracDisorder} are being explored. New developments to diagnose critical behaviour in the Hubbard model using neural networks have been considered~\cite{Neural}, as well as, engineering of bound states via disorder~\cite{BoundDisorder}. 
Some standard schemes used to study the referred phenomenology are approximations to the extended and hybrid Bose-Hubbard model using G\"utzwiller ansatz~\cite{GWA}, Stochastic Mean-Field Theory~\cite{Lewenstein,SMFT13,Yukalov}, Self-energy Functional Theory~\cite{SelfEnergy} and the use of Monte-Carlo simulations~\cite{PolletReview,Pollet2}. Numerical exact diagonalization simulations in the full quantum limit are difficult to perform since the size of Hilbert space grows exponentially with the number of lattice sites. Moreover, Renormalization Group schemes are computationally very expensive for systems in dimensions larger than one. Additionally, in the weakly interacting regime Monte-Carlo simulations show strong finite size effects~\cite{Capogrosso}. This makes difficult the characterisation of the states in the simulations and the identification of the corresponding quantum many-body phases of matter. Therefore, other alternatives to address the problem in the lattice with many sites are desirable. A suitable approximation scheme is given by using the Gross-Pitaevskii (GP) equation to effectively describe the ground state of weakly interacting bosonic systems at low temperatures~\cite{Pitaevskii-Stringari,Fallani}. The limitations of this approach are that fluctuations in the number of particles are always poissonian in GP calculations, unable to represent MI states or ensembles of them while depletion is neglected. However, this mean field treatment allows to investigate either stationary or dynamical transport properties in the superfluid (SF) regime (compressible phases) away from insulating states driven by interactions (incompressible phases). The aim of our study is to give an effective picture of the role of geometry in the transport properties of the quantum degenerate gas in the limit of weak interactions in an optical lattice. To this end, we employ an approximate treatment using a mean-field description, GP picture. In this work, we restrict our attention to the characterisation of disorder induced phenomenology and the interplay with the lattice topology via numerical experiments.

In order to identify different signatures of localisation, we describe the weakly interacting bosonic gas in the limit of low temperatures ($T=0$), where the system is accurately described by a macroscopic wave function $\psi({\mathbf x},{\mathbf y},t)$ describing the condensate component of the system\cite{LesHouches, Pitaevskii-Stringari}. The time evolution of the system is governed by the GP equation under the confinement of the optical lattice. This equation allows us to model the stationary states of the system subject to different lattice geometries, disorder strength $\delta$, and effective weak on-site interactions. The analysis of our simulations allows us to identify localisation features by varying  the disorder strength ($\delta$) and the system size ($\Omega$). The number of minima induced by the optical lattice where the atoms localise having maximum density peaks defines precisely such a system size. In the absence of disorder ($\delta=0$), the system is a quantum fluid with translational invariance. As the disorder amplitude $\delta$ is increased, the formation of regions where the density is negligible appear, that is, a continuous density of the condensate is replaced  by islands across the 2D lattice. This generates disconnected islands that lead to the suppression of atomic transport across the whole lattice. We arrived to this conclusion after performing a statistical analysis of several quantities, numerically determined, for the condensate wavefunction confined in the 2D disordered lattices. In particular, we studied the average of peak heights as well as the  density across the lattice. Fragmentation of the condensate in islands is qualitatively similar to depletion in the sense that a fraction of the condensate extinguishes in certain regions. However, since we are only considering the ground-state of the condensate, that is the  $T=0$ wave function, we are unable to consider finite $T$ effects\cite{Falco,Yukalov}. As found in the literature, the inhomogeneity introduced by a random potential generates excitations of low energy finite momentum modes\cite{Lugan,SanchezPalencia0, Muller,Muller2}, or causes that a fraction of the superfluid become a normal fluid in the 3D case~\cite{Huang}. It has been found in experiments with optical lattices that depletion affects marginally ($\lesssim10\%$) the behaviour of the atomic density in low dimensions (1D or 2D) when optical lattice depths are  lower than those considered in this work\cite{KetterleDepletion}. This supports the use of an effective description given solely in terms of the behaviour of the condensate component in the system as a first approximation. In ultracold gases with weak interactions, this suggests that disorder leads to the fragmentation of the SF in the system. This is the well known scenario that leads to the formation of a Bose-Glass for larger values of disorder strength, which is also a fragmented SF state\cite{Fallani,BGfrag}. However, it is important to stress that in our study we considered both, effective interactions and disorder strengths for which the system remains in the superfluid phase, away from the Bose glass and insulating phases.

This work is organised as follows. In the section model, we describe the system and parameters used in our simulations to analyse localisation effects. In section Disorder induced localised states in 2D lattices, we characterise the localisation or density fragmentation, by performing a robust numerical analysis of several observables. Then in section Ground state energy spectra of localised states we study the role of disorder in the 2D lattices in the momentum space. Finally in the last section, we summarise our findings.

\section*{Model: Weakly Interacting ultracold atoms in 2D disordered lattices}

\begin{figure}[t]
\begin{center}
\includegraphics[width=3.5in]{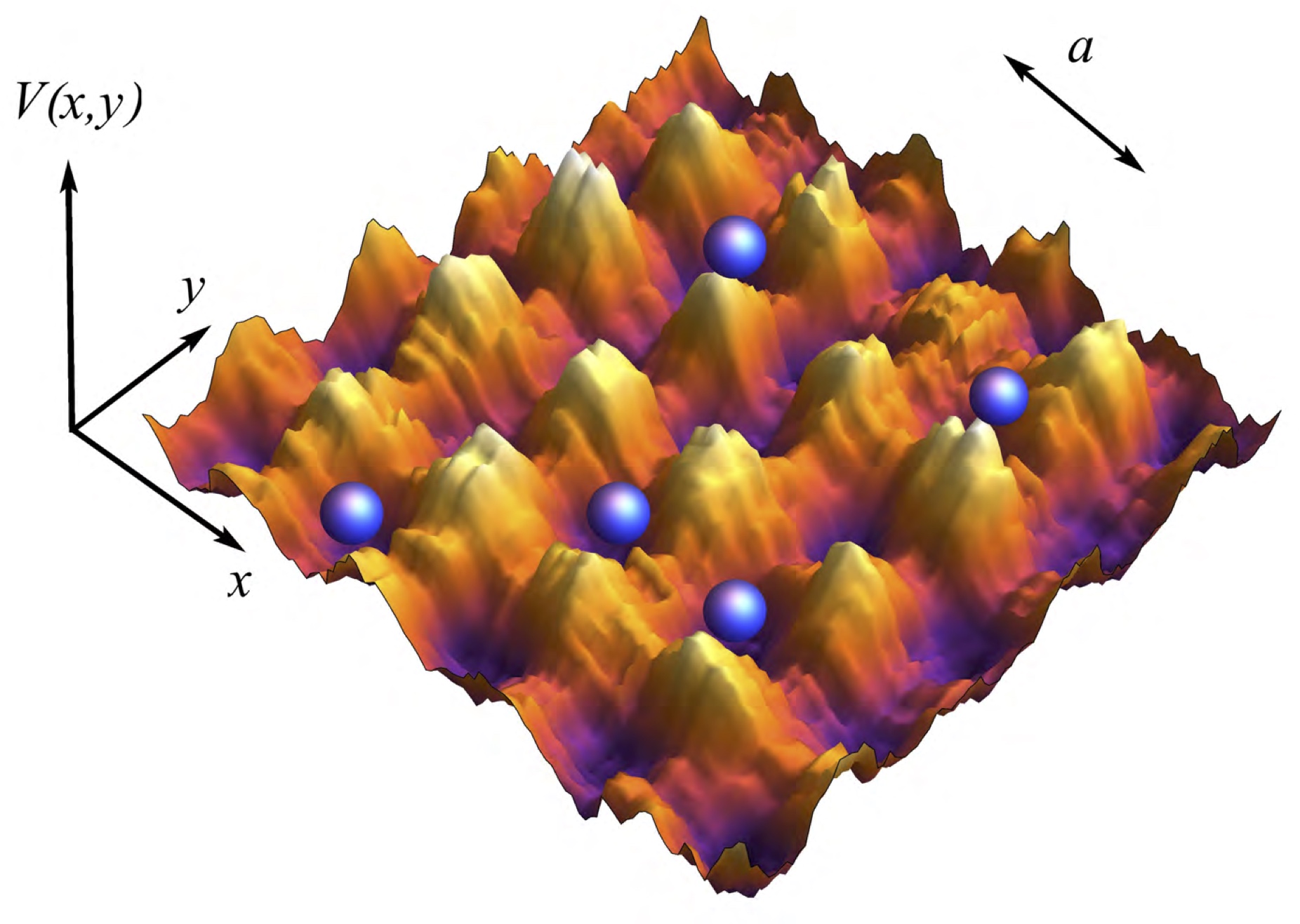} 
\end{center}
\caption{Schematic representation of the two dimensional square disordered lattice potential.}
\label{Fig1}
\end{figure}

To analyse the influence that disorder and the lattice geometry have in producing localised states in ultracold systems, we consider a weakly interacting gas of Bose atoms confined in several two dimensional lattices, subjected to white noise disorder of variable strength. The system is described by the mean field Gross-Pitaevskii equation for the amplitude of the wave function $\psi$,
\begin{equation}
i\hbar\frac{\partial \psi}{\partial t}=-\frac{\hbar^2}{2m}\nabla_\perp^2 \psi+\left (V^\delta_{\textrm{Latt}}(x,y) +g_{\mathrm{2D}}|\psi|^2 \right)\psi,
\label{GPE}
\end{equation}
where $\nabla_\perp^2=\frac{\partial^2}{\partial x^2}+\frac{\partial^2}{\partial y^2}$, and $g_{\mathrm{2D}}$ being the effective coupling interaction in 2D \cite{RefU2D, U2D, Salasnich, Mateo, Bao, Trallero}. The potential $V^{\delta}_{\textrm{Latt}}(x,y)$ is given by a two dimensional lattice with triangular, square or honeycomb symmetries \cite{LUIS,LUIS2},
\begin{eqnarray*} \nonumber
V^\delta_{\hexagon}(x,y)&=& V_{0}^\delta \Bigg[ \cos \left( \frac{4\pi y}{3a} \right)+ \cos \left({\frac{2\pi x}{\sqrt{3}a}-\frac{2 \pi y}{3a}}\right) + \\
&+&\cos \left(\frac{2\pi x}{\sqrt{3}a}+ \frac{2 \pi y}{3a}\right) \Bigg],
\\
V^\delta_{\square}(x,y)& =& V_{0}^\delta   \Bigg[  \sin^2 \left({\frac{\pi x}{a}}\right)+  \sin^2 \left({\frac{\pi y}{a}}\right) \Bigg],\\
V^\delta_{\triangle}(x,y)&=& V_{0}^\delta  \Bigg[ \cos \left( \frac{\pi}{4} -\frac{4\pi}{3} \frac{y}{a'} \right) + \\
&+&\cos \left( \frac{\pi}{4}+\frac{4\pi}{3} \frac{y}{2a'} \right)  \cos \left(   \sqrt{3}\frac{4\pi}{3} \frac {x}{2a'} \right)\Bigg],
\label{potentials}
\end{eqnarray*} 
with $a$ the lattice constant and $a'=0.55a$,  labels $\{\hexagon,\square,{\triangle}\}$ identify honeycomb ($z=3$), square ($z=4$) and triangular lattices ($z=6$), with $z$ being the coordination number (the number of nearest neighbours), $V_{0}^\delta= V_{0}^{\mathrm{Latt}}(1+{\epsilon_\delta(x,y)})$, being $V_0^{\hexagon}=2.66V_0/12$, $V_0^{\square}=V_0/2$, and $V_0^{\triangle}=3.45V_0/12$. The function $\epsilon_\delta(x,y)$, representing a non correlated disorder, takes random values varying in the interval $[-\delta,\delta]$ at each point $(x,y)$ being $\delta$ scaled in the same energy units as $V_0$. Thus, the total potential depth at each point $V_{\textrm{Latt}}^\delta(x,y)$ (disordered contribution and optical lattice) is the result of adding/subtracting a random number $\epsilon_\delta$ to the amplitude of the potential $V_{\textrm{Latt}}^{\delta=0}(x,y)$ at each point $(x,y)$. To perform a reliable analysis of the physical quantities and have meaningful predictions, we average over an ensemble of realisations for each value of the disorder amplitude $\delta$. In Fig. \ref{Fig1} we show a fragment of the energy landscape used in our study, illustrating a particular realisation of disorder in a square geometry. Similar to real experiments, multiple realisations will be considered, as one random realisation fails to represent the typical behaviour of multiple scattering events due to uncorrelated disorder. The way in which the disorder has been simulated, warrants that, although the lattice symmetry is altered, the underlying structure is preserved. It is important to note, that this way of setting disorder through the whole continuous coordinate space where the lattice is defined, resembles analog real systems in condensed matter where each particle ``feels" a different energy landscape associated to the disorder in the otherwise perfect lattice. Additionally, this way of implementing disorder is similar to consider a random energy shift at each lattice site, as originally formulated~\cite{Anderson}. The disordered potentials experimentally created are the result of either, the light arising from an optical speckle field produced when a laser beam passes through a diffusing plate \cite{Clement,Sanchez-Palencia,Plodzien,Piraud}, or the combination of both, a standing wave with a defined lattice geometry and the light arising from an optical speckle field \cite{White}.

\subsection*{Simulation considerations and experimental parameters}

In our simulations, we consider typical parameters of ultracold  $^{87}$Rb atoms \cite{Bloch} confined in an optical lattice potential in 2D, having depths of $V_{0}\sim12 E_R$ and a lattice spacing $a=532$ nm, where $E_R$ is the recoil energy [$E_R= h^2/(8 m a^2)$]\cite{RMP}. At these optical lattice depths the single band Bose Hubbard model\cite{RMP,Fisher,Lewenstein} is an accurate description. Other natural scales for the energy are the hopping amplitude between nearest neighbours $t_0$, and the on-site interaction strength in the optical lattice $U_{\mathrm{2D}}$ \cite{U2D-BH}. For the lattices here analyzed, the tunneling amplitude $t_0$ can be estimated analytically in the case of the square lattice \cite{RMP}, and numerically in the cases of triangular and honeycomb lattices \cite{Walters,ModugnoPRA,Vanderbuilt,Lee}. As it is well known, the appropriate parameter to verify the range of validity of the regime away from strong interaction effects, is the quotient between $t_0$ and $U_{\mathrm{2D}}$. As reported in the literature, for a filling factor of $n=1$, the SF-MI transition occurs for $\tilde{t}_c^{\mathrm{\square}}=t_0/U_{\mathrm{2D}}\approx0.06$, $\tilde{t}_c^{\mathrm{\triangle}}=t_0/U_{\mathrm{2D}}\approx0.04$ and $\tilde{t}_c^{\mathrm{\hexagon}}=t_0/U_{\mathrm{2D}}\approx0.08$, for square\cite{Porto}, triangular and hexagonal\cite{Teichmann} lattices respectively. Therefore, stronger effective on-site interactions than those bounds in each case are needed to be able to represent sub-poissonian on-site number fluctuations, that the GP model is unable to reproduce.  In our calculations we use $U_{\mathrm{2D}}/E_R\sim0.01$. In this case, the system is very far away from the interaction driven insulating states. Therefore, in our simulations there is no competition with Mott states. Moreover, we neglect quantum depletion effects as we are in the $T=0$ regime\cite{KetterleDepletion}. Therefore, as we are deep in the SF regime in 2D, the GP theory is a pertinent description in the presence of disorder.

\begin{figure}[t]
\begin{center}
\includegraphics[width=3.5in]{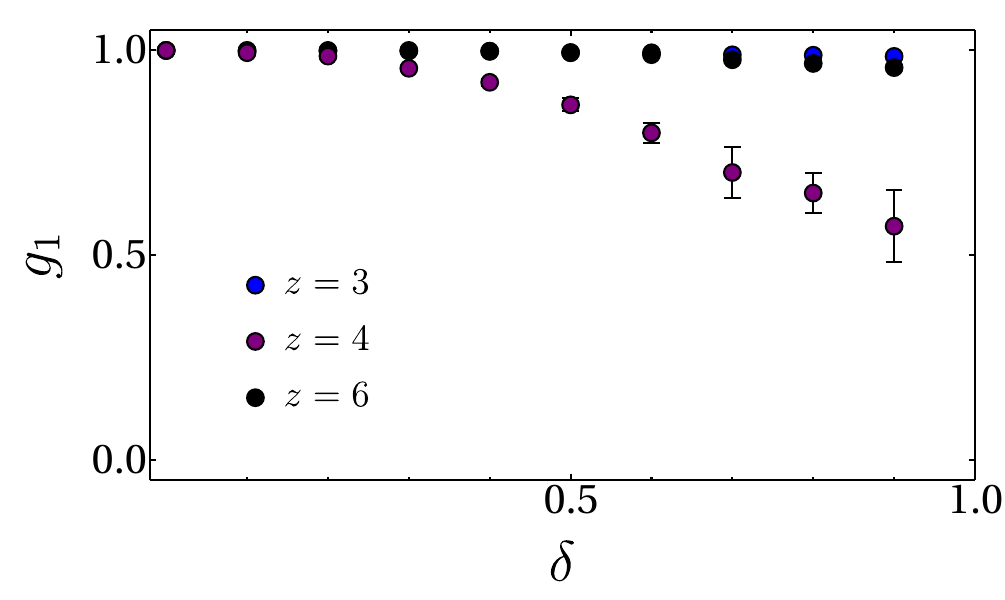} 
\end{center}
\caption{Correlation function between nearest neighbours $g_1$ as a function of $\delta$ for the hexagonal ($z=3$), square ($z=4$) and triangular ($z=6$) lattices. Error bars in this figure are associated to the ensemble of 50 realisations for each value of $\delta$.}
\label{FigPC}
\end{figure}

A quantity that allows us to establish the range where disorder amplitude can be varied before the system enters into the phase in which the density of the condensate becomes discontinuous across the lattice, is the correlation function at the nearest neighbor distance, $g_1$. This quantity can be defined in analogy to the so called phase coherence $\chi(t)$ that measures the correlation between values of the Bose condensate wave function at points separated one lattice constant $d$. The phase coherence defined as \cite{Nesi-Modugno} $\chi(t) \equiv  \left |\int d^3x  \> \psi^* (r,z; t) \psi(r, z+d; t)  \right |^2$, provides reliable information of the degree of coherence of the condensate inside the optical lattice\cite{Nesi-Modugno}. In our case, the correlation function between nearest neighbors, is defined as $$g_1= \frac{1}{z^2} \sum_{m \in n.n.} \left |\int dx dy \> \psi^* (x,y) \psi(x+x_m, y+y_m)  \right |^2$$
where $z$ is the coordination number and $m$ is a label that identifies the nearest neighbors sites with respect a site located in $(x,y)$ position. For the particular case of the square lattice we have $(x_m/a,y_m/a)= \{ (0,1), (0,-1), (1,0), (-1,0) \}$, while being $(x_m/a,y_m/a)=\{ (1,0), (1, \sqrt{3}/2), (-1/2, \sqrt{3}/2), (-1,0), (-1/2,-\sqrt{3}/2), (1/2, -\sqrt{3}/2) \}$ for the triangular lattice, and $(x_m/a,y_m/a)= \pm\{  (1,0), (-1/2, \sqrt{3}/2), (-1/2, -\sqrt{3}/2) \}$ for the honeycomb geometry. The correlation $g_1$ gets lost when the density of the condensate is discontinuous, thus preventing the GP approach. Therefore, besides considering values of effective interaction for which the system is well below the MI phase, we shall also consider values of disorder strength that guaranty that the interacting Bose gas remains sufficiently smooth, in order to ensure that our description of the condensate in terms of solutions of a differential equation still makes sense. We should mention here that the phase coherence is a quantity that allows to distinguish when the condensate is not a Bose glass \cite{SMFT13}. In Fig. \ref{FigPC} we show the behaviour of $g_1$ as a function of $\delta$ for the analysed lattices. The values of the correlation $g_1$ in this plot are normalised with respect to its value at zero disorder. Error bars in this figure are associated to the ensemble of realisations for each value of $\delta$ (measured in units of recoil energy). As can be appreciated from Fig. \ref{FigPC}, for values of $\delta$ below 1, the system remains distributed across the lattice. Thus we can safely consider disorder strengths below one hundred percent of the recoil energy $E_R$ to be in the superfluid regime.

In typical ultracold experiments besides disorder in the lattice, the atoms move under the influence of a harmonic confinement, so that $V \to V+V_T$, with  $V_T=\frac{1}{2}m \omega^2 \left(x^2+y^2\right)$. Although this contribution can also be considered in our mean field analysis, and in fact we performed calculations to investigate the modifications that the inhomogeneity introduces, we note that real analog systems are not under the presence of a harmonic potential, but instead each particle is immersed in a very large (infinite) disordered lattice. Also, near the center of the trap in a typical ultracold atom experiment it is reasonable to neglect harmonic confinement for sufficiently large lattices. Thus, our analysis will focus on lattices without the harmonic confinement for simplicity. 

\section*{Disorder induced localised states in 2D lattices}
Using the effective model given by Eq.(\ref{GPE}), we study the emergence of localisation for honeycomb, square and triangular lattices as a function of the disorder amplitude and the system size. In order to describe the ground state of the system, we numerically solve Eq. (\ref{GPE}) using imaginary time evolution ($t\to i\tau$), which is equivalent to finding the lowest energy solution or steady-state solution $\psi\to\psi_0$\cite{fn2}. This solution represents the amplitude of the macroscopic wave function of the system at $T=0$ in the lowest energy state. We simulate the GP equation in continuous coordinate space for a given disorder amplitude $\delta$ and analyse the obtained density profiles $\rho(x,y)=| \psi_0(x,y)|^2$. As explained previously, the limitations of our treatment are that it does not account for depletion of the condensate, non-poissonian density distributions and it is a zero temperature model. However, it allows to study the combined effects of geometry and disorder, which is the main purpose of this work. The identification of localisation is made from the stationary state by first quantifying the density across the lattice sites, as a function of disorder for a given lattice geometry. Then, to further comprehend the emergence of localisation, we study the behaviour of the stationary localised states as a function of system size. For our calculations, we consider lattices having number of sites used in typical experiments performed in 2D lattices. In order to have meaningful quantities as a function of the disorder amplitude, we consider sets of $\sim 50$ realisations of random numbers for each value $\delta$ and a given lattice size $\Omega$. The size $\Omega$, is identified as the number of minima of the potential $V^{\delta}_{\textrm{Latt}}(x,y)$ without disorder. The initial state used in all of our calculations to reach the steady state is a constant distribution $\psi_{0}(x,y)=\textrm{constant}$ \cite{fn2}. 

\begin{figure}[t]
\begin{center}
\includegraphics[width=3.7in]{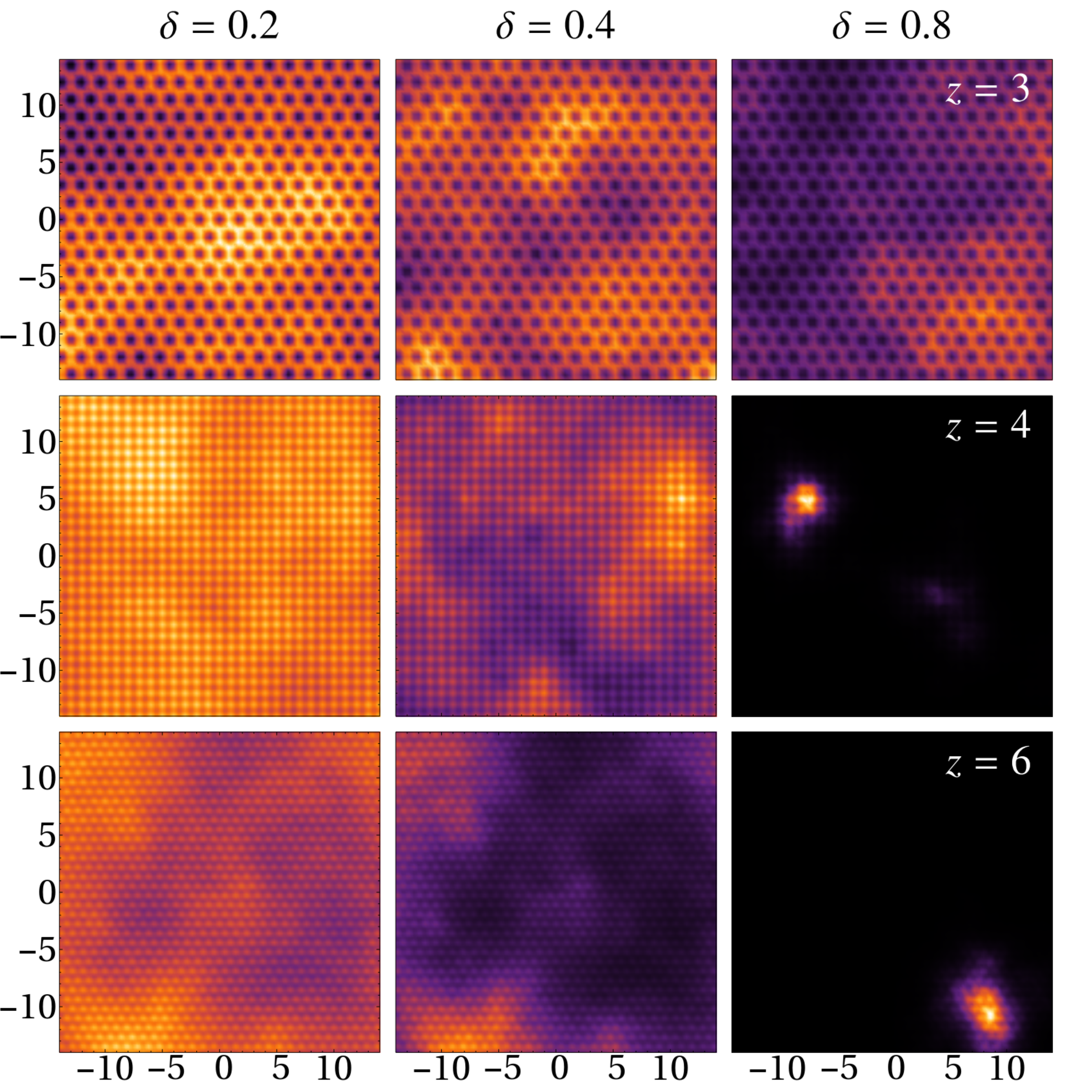}
\end{center}
\caption{Fragments of the density profiles for honeycomb, square and triangular lattices with a disorder amplitude $\delta=0.2, 0.4, 0.8$ with lattice sizes $\Omega\sim10^3$. Parameters are: $V_0 = 12 E_R, U = 0.01 E_R$, spacing is in units of the lattice constant $a$. The bright regions correspond to density maxima while dark regions absence of SF.}
\label{Fig2}
\end{figure}

The information obtained directly from our numerical calculations can be summarised as follows. 1. At zero disorder strength, gaussian peaks centred at lattice sites fill the entire space, being the amplitude of those peaks the same across the lattice, except at the edges where the boundary condition (end of the lattice) give rise to peaks with lower amplitudes, 2. When the magnitude of disorder $\delta$ is non zero, gaussian peaks are not distributed uniformly in the whole lattice, instead of that, several sites in the lattice show a diminished density, thus exhibiting peaks with lower amplitudes. We fixed an arbitrary criterium to count the non negligible peaks contributing to the superfluid density (for a given disorder strength, only peaks having amplitudes larger than 5 percent of the highest amplitude are considered for the statistics), 3. As expected, different realisations of disorder associated to a given value of the disorder amplitude, produce different distributions of peaks across the lattice. What it is important to stress is that different realisations of disorder have, on average, the same distribution of heights and density peaks, 4. As the size of the disorder amplitude increases, the spatially distributed random peaks become sparse, and therefore, in accordance with the condition of constant density, the heights of the peaks become taller, and 5. As the magnitude of disorder is increased, the zero disorder scenario is replaced by a fragmented  density across the lattice. For illustration purposes in Fig. \ref{Fig2}, we show the stationary density profiles for one of the realisations and different lattice geometries ($z=3,4,6$). These profiles correspond to disorder amplitudes of $20\%$, $40\%$ and $80\%$ of the value of $V(x,y)^{\mathrm{lattice}}$ at each point $(x,y)$. The analysis of many realisations of these density distributions is the core of our work.  

In the typical model of localisation induced by disorder without interactions, the envelope of the wave function ($\phi$) gets localised exponentially\cite{Anderson}. In 1D, in an isotropic medium with randomly distributed scatterers $\phi(x)\sim\exp(-|x-x_0|/(\xi_{\mathrm{Loc}})$ where $x_0$ is some arbitrary point in the space, $\xi_{\mathrm{Loc}}$ is a localisation length\cite{Volthardt}. Within the self consistent theory \cite{WolfeRev},  the localisation length depends on disorder as $\xi_{\mathrm{Loc}}\sim l\exp(kl)$ in 2D and $\xi_{\mathrm{Loc}}\sim l$ in 1D, where $l$ is the scattering mean free path and $k$ the wave momentum. In low dimensions (1D and 2D) in the absence of interactions, the localisation length diverges for weak disorder strength, which means that there is no finite value of disorder for which the system can be an ideal conductor. Almost ten years ago, experiments performed in atoms of $^{87}$Rb confined in one dimension, in a quasi-periodic lattice \cite{Roati}, and in a weak disordered optical potential \cite{Sanchez-Palencia}, showed the typical exponentially localised density profile behaviour. In our problem, instead of studying such a long tail exponential behaviour, we focus our attention on the short distance density variations to characterise the localisation phenomenon. The effects of disorder are captured by performing a robust study of the density profiles. In particular, taking into account the information obtained from our numerical calculations, the quantification of disorder effects in our system will be established in terms of two observables, the peaks heights distribution $h(\delta)$ and the peak fraction $p_f$, as a function of disorder amplitude $\delta$. Here we stress that the observables to be studied can be experimentally accessed in actual ultracold atom experiments.
     
\begin{figure}[t]
\begin{center}
\includegraphics[width=0.48\textwidth]{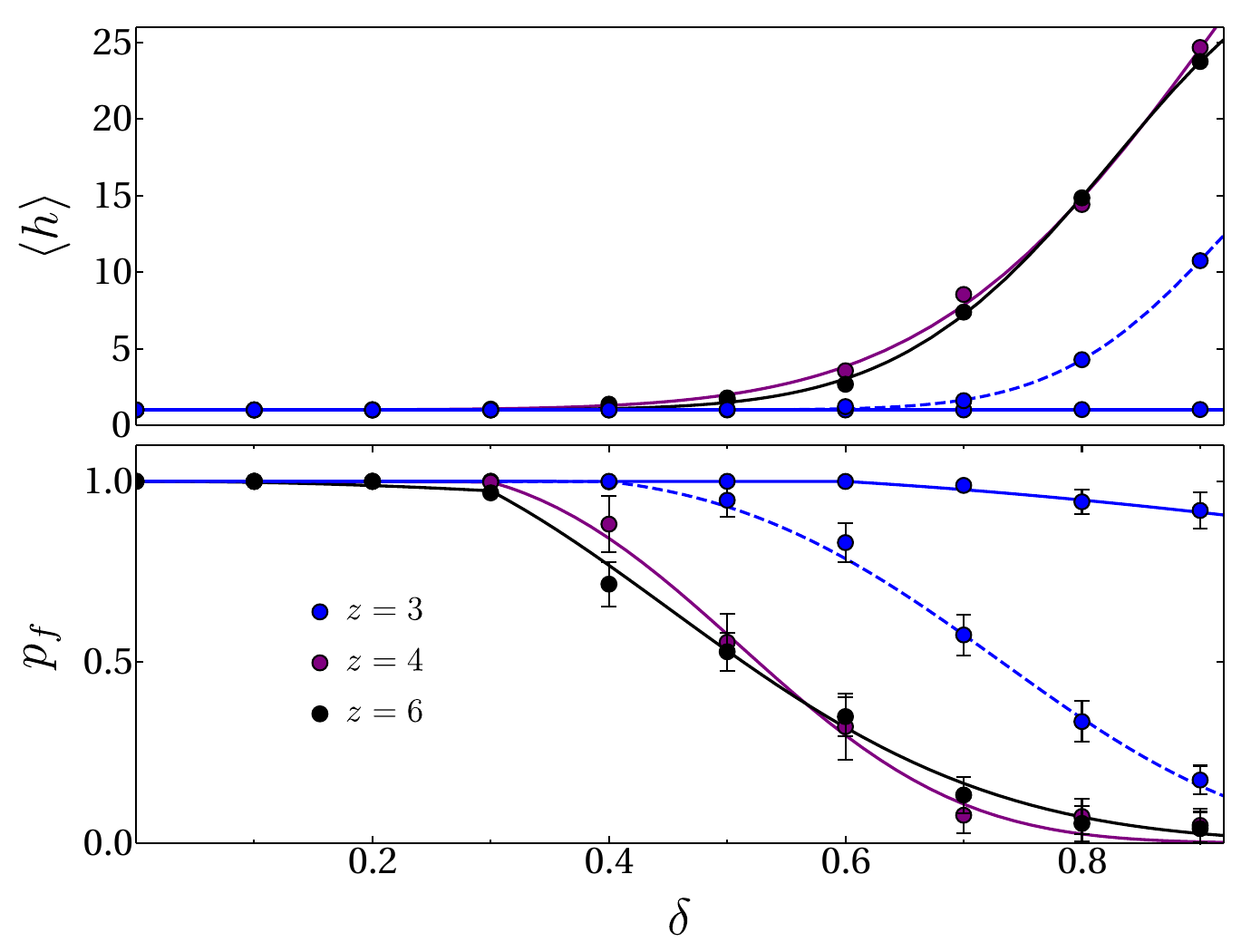}
\end{center}
\caption{Average of the peaks heights $\langle h \rangle$ (top) and the peak fraction $p_f$ (bottom) as a function of disorder $\delta$ in units of  $E_R$. $\langle h \rangle$ is normalised to the largest height amplitude. Blue, purple and black symbols correspond to honeycomb ($z=3$), square ($z=4$) and triangular ($z=6$) lattices respectively. Parameters are: $\Omega\sim10^3$, $V_0=12 E_R$ (solid lines) and $V_0=24E_R$ (dashed line for honeycomb), $U=0.01E_R$, with $\sim 50$ disorder realisations at each data point, the bars denote the standard deviation for each ensemble of disorder amplitude $\delta$. The honeycomb lattice with $V_0=24E_R$ (dashed blue) and the square lattice with $V_0=12E_R$ (solid purple) have equivalent tunneling amplitudes in the tight binding limit, see the main text for details. The continuous lines for $\langle h\rangle$ are given for triangular and square by: $1+\exp(b_0+b_1\delta+b_2\delta^2)$, with parameters (triangular, square): 
$b_{0,\textrm{Latt}}=(-12.53,-7.84)$, 
$b_{1,\textrm{Latt}}=(31.42,19.98)$, and 
$b_{2,\textrm{Latt}}=(-15.58,-8.61)$.
 For the honeycomb lattice, $\langle h\rangle\approx 1$ for $\delta\lesssim1$ with $V_0=12 E_R$; for $V_0=24E_R$,
$b_{0,\hexagon}=-26.04$,
$b_{1,\hexagon}=54.58$, and 
$b_{2,\hexagon}=-25.68$. 
The solid lines for $p_f$ are given by: $1-\alpha\delta^2$ for $\delta\lesssim\delta_c$ and $A \exp(\beta_1\delta+\beta_2\delta^2+\beta_3\delta^3+\beta_4\delta^4)$ for $\delta>\delta_c$.
The fitting parameters are for $V_0=12E_R$ (triangular, square, honeycomb):
$\alpha_{\textrm{Latt}}=(0.29,0.022,0.00051)$, 
$\delta_{c,\textrm{Latt}}\approx (0.3,0.3,0.6)$, 
$A_{\textrm{Latt}}=(1.023,0.98,0.99)$, 
$\beta_{1,\textrm{Latt}}=(1.26,0.0,0.21)$, 
$\beta_{2,\textrm{Latt}}=(-4.17,3.59,-0.34)$, 
$\beta_{3,\textrm{Latt}}=(-1.93,-11.52,0.005)$, 
$\beta_{4,\textrm{Latt}}=0.0$. 
For the honeycomb lattice with $V_0=24E_R$, 
$\alpha_{\hexagon}=0.006$, 
$\delta_{c,\hexagon}\approx 0.4$, 
$A_{\hexagon}=0.98$, 
$\beta_{1,\hexagon}=0$, 
$\beta_{2,\hexagon}=0.69$, 
$\beta_{3,\hexagon}=0$,
$\beta_{4,\hexagon}=-3.64$.
}
\label{Fig3}
\end{figure}

To investigate the dependence of $h$ and $p_f$ on $\delta$ we proceed as follows. For fixed disorder strength $\delta$ and lattice size $\Omega \sim 10^3$, we performed $\sim$50 realisations of random numbers. From each density profile of the condensate, we analyse the amplitude of the peaks at each lattice site by selecting the non-negligible peaks, using the criteria described above. With this knowledge we determine both, the relative heights and the peak fraction for each realisation. Then, we take the average of $h$ over the ensemble. In Fig. \ref{Fig3}, we summarise the results obtained for honeycomb, square and triangular lattices. There, we plot the average of the peaks heights $\langle h \rangle$ (top) and the peak fraction $p_f$ (bottom) as a function of disorder $\delta$ in units of the recoil energy $E_R$. $\langle h \rangle$ is referenced with respect to the average heigh value without disorder. Blue, purple and black symbols correspond to honeycomb ($z=3$), square ($z=4$) and triangular ($z=6$) lattices respectively. As can be appreciated from Fig. \ref{Fig3}, square and triangular lattices suggest a smooth transition to localisation for $\delta\gtrsim 0.3$ where both, $\langle h \rangle$ and $p_f$ start to change. In other words, the density collapses into a single density peak randomly located across the lattice. In contrast, localisation is not observed for the honeycomb lattice for the same range of disorder strength. Interestingly, the continuous lines in Fig. \ref{Fig3}, fitting the data of each geometry, allow to conclude that there is not sensitivity of the localisation appearance on the lattices having coordinations $z=4$ and $z=6$. Blue dashed lines in the top and bottom figures of Fig. \ref{Fig3} correspond to the honeycomb lattice, but with a different value of the potential depth with respect to that considered for the continuous lines. While continuous lines correspond to $V_0/E_R\sim 12$, dashed lines label $V_0/E_R\sim 24$. Therefore, from this figure one can conclude that the stronger tunneling energy of the honeycomb lattice against the lower for square and triangular lattices, for the same lattice amplitude, plays a determinant role in the occurrence of localisation.
 
\begin{figure}[t]
\begin{center}
\includegraphics[width=0.58\textwidth]{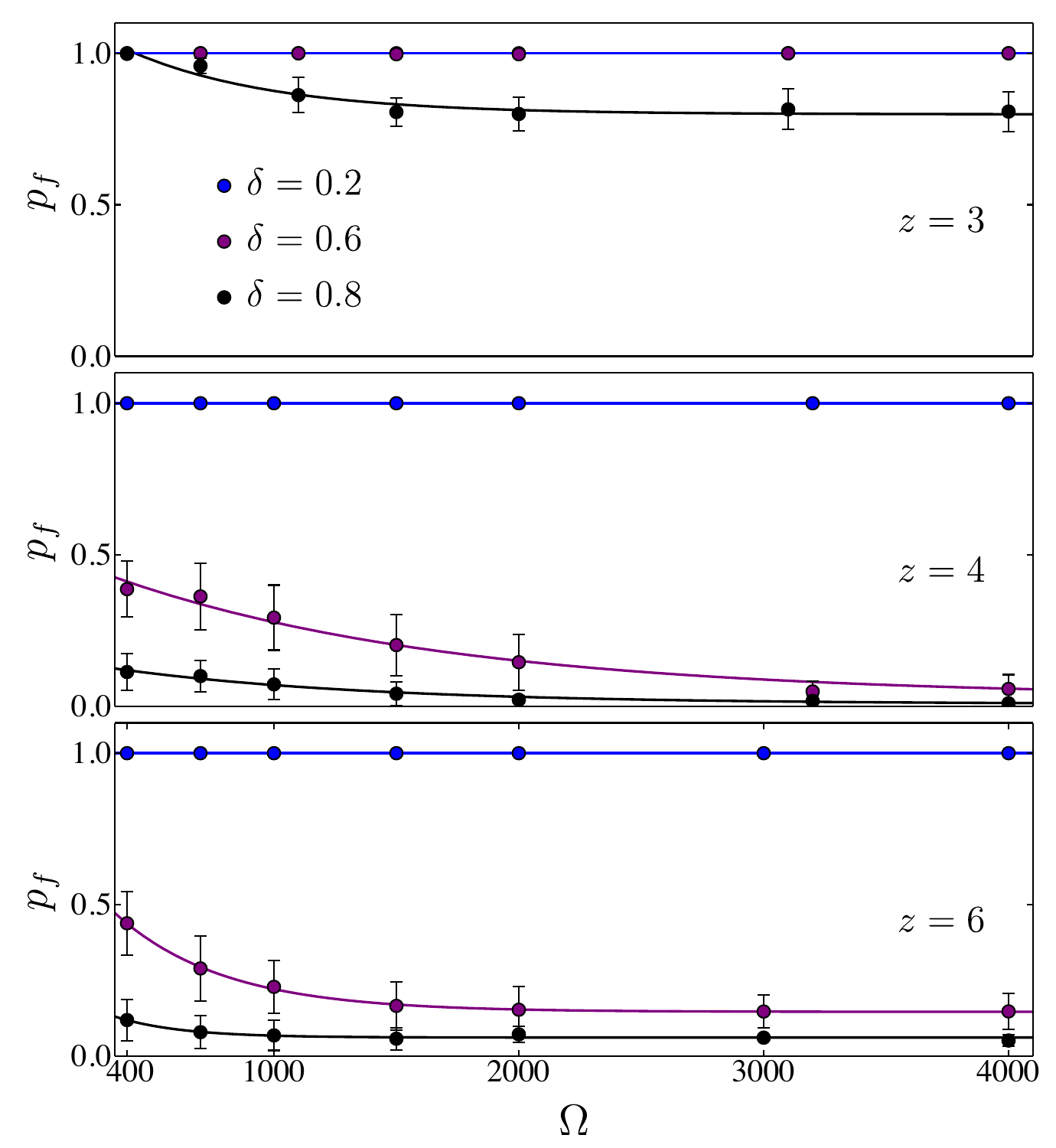}
\end{center}
\caption{Normalized fraction of peaks as a function of the lattice size $\Omega$. The values of the disorder amplitude are $\delta=20\%\textrm{(blue)},60\%\textrm{(purple)}, 80\%\textrm{(black)}$ of the potential depth $V_0=12 E_R$. From top to bottom panels correspond to honeycomb, square and triangular lattices respectively. Interestingly, the honeycomb lattice  ($z=3$) is resilient towards disorder induced localisation, as compared to other geometries with more nearest neighbours ($z=4,6$).
The fitted lines different from $p_f\sim1$ are given by: $p_f\sim B\exp(-c \Omega)+D$. For honeycomb: $\delta=0.8$, $(B,D,c)=(0.42,0.8,0.0017)$. For square: $\delta=0.6$, $(0.51,0.03,0.00073)$; $\delta=0.8$, $(0.16,0.0072,0.00097)$. For triangular: $\delta=0.8$, $(0.72,0.15,0.0023)$; $\delta=0.8$, $(0.26,0.06,0.0037)$.
}
\label{Fig4}
\end{figure}

We also investigate finite size effects considering lattices of size $ 5 \times10^2 \lesssim \Omega \lesssim 5 \times 10^3$. In particular, we study how the density fragmentation of the condensate component occurs as disorder amplitude is increased and the system size is varied. In Fig. \ref{Fig4}, we plot the fraction of peaks as a function of $\Omega$ for several values of $\delta$, blue, purple and black points correspond to $\delta=20\%$, $60\%$ and $80\%$ respectively, panels from top to down correspond to honeycomb, square and triangular lattices respectively. Several remarks can be extracted from this figure. First, in our simulations we observe that for disorder amplitudes of $\sim40 \%$ the peak fraction never reaches a constant behaviour for triangular and square geometries. In the region $30\%\lesssim\delta\lesssim50\%$ we observe that large fluctuations in $p_f$ occur preventing stabilisation. Although this behaviour is similar that observed in typical phase transitions in which critical slowing down and enhanced fluctuations occurs\cite{SlowD}, we should stress that the treatment used for our study is unable of describing phase transitions. For $\delta\gtrsim 50\%$, (in figure  \ref{Fig4} we just illustrate  $\delta = 60 \%$ and $80 \%$) the peak fraction saturates to a small but constant value for these geometries. That is, we see that the fraction of peaks becomes independent of the lattice size for $\Omega > 3 \times 10^3$ for $\delta\gtrsim 50 \%$. This result confirms that lattices with coordinations $z=4$ and $z=6$ suffer a noticeable density fragmentation for disorder strengths above $\delta\gtrsim 30 \%$. Although, we observe that the system does not reach the limit where $p_f\to 0$, as in the standard notion of a single localised peak, this result for our lattices in 2D is analogous to the 1D case in which, for non-trapped gases and finite repulsive interactions, the Bose gas populates a finite number of localised single-particle Lifshitz states\cite{BGfrag}. In agreement with the results found for lattices of size $\Omega \sim 10^3$, honeycomb lattices do not exhibit such fragmentation for the same disorder strengths and the same lattice amplitude. It is important to stress that this   density fragmentation observed for the analysed lattices is away of the Bose-Glass phase\cite{Fallani,BGfrag,Giamarchi,Laflorencie} since, as described in the previous section, the correlation function at the nearest neighbour distance is well above of zero, thus indicating that the system is not an insulator.

\section*{Ground state energy spectra of localised states}
Complementing our analysis, we investigate the energy spectrum of the stationary states from our simulations. To proceed, we consider the energy associated to the stationary GP equation as a function of the disorder strength $\delta$ and a fixed value of the mean field interaction $U$. The energy functional is given in terms of the stationary state $ \psi_{0}=\lim_{t\to\infty}\psi=\lim_{t\to 0}\psi$ by,
\begin{equation}
E_0= \int d^2r \left \{ \psi_{0}^*\left[ -\frac{\hbar^2}{2m}  \nabla_\perp^2   +V^\delta_{\textrm{Latt}}(x,y) \right]  \psi_{0} + \frac{U}{2}  | \psi_{0}|^4 \right \},
\end{equation}
where the disorder is considered in the effective potential $V^\delta_{\textrm{Latt}}$ for  honeycomb ($V^\delta_{\hexagon}$), square ($V^\delta_{\square}$) and triangular ($V^\delta_{\triangle}$)  lattices (see section Model). The integrand is the ground state energy density in continuous coordinate space. To find the ground state energy density in the reciprocal space, namely the ground state energy spectra, we first rewrite the kinetic term in a quadratic form,  
\begin{equation}
\int d^2 r \psi_{0}^* \left( -\frac{\hbar^2}{2m}   \nabla_\perp^2 \right) \psi_{0} =   \frac{\hbar^2}{2m} \int d^2r\nabla_\perp \psi_0^* \cdot \nabla_\perp \psi_0.
\end{equation}
Then, it follows from the use Parseval's theorem\cite{Nore}, the expression in momentum space,
\begin{equation}
\epsilon(k_x,k_y)=\frac{1}{2 m}\mathbf{P}_{\psi_0}^*\cdot \mathbf{P}_{\psi_0}^{\phantom{*}}+ \frac{|U|}{2} \left | \int d^2 r  e^{i \bf{k}_\perp \cdot \bf{r}_\perp} |\psi_{0}|^2\right |^2
+\left | \int d^2 r  e^{i \bf{k_\perp} \cdot \bf{r_\perp}}  {\sqrt {V^\delta_{\textrm{Latt}}(x,y)}}  \psi_{0}\right |^2,\quad \mathbf{P}_{\psi_0}^{\phantom{*}}=\hbar \int d^2 r e^{i \bf{k_\perp} \cdot \bf{r_\perp}}  \nabla_\perp {\psi_{0}},
\label{E_spectrum}
\end{equation}
with ${\bf{r}}_\perp=x\hat e_x+y\hat e_y$,  ${\bf{k}}_\perp=k_x\hat e_{x}+k_y\hat e_{y}$ and $\hat e_{x/y}$ unit vectors in 2D.
As it turns out from Eq. (\ref{E_spectrum}), the ground state energy spectrum associated to a particular realisation of disorder encodes the structure of the ground state energy in momentum space. 

\begin{figure}[t]
\begin{center}
\includegraphics[width=0.48\textwidth]{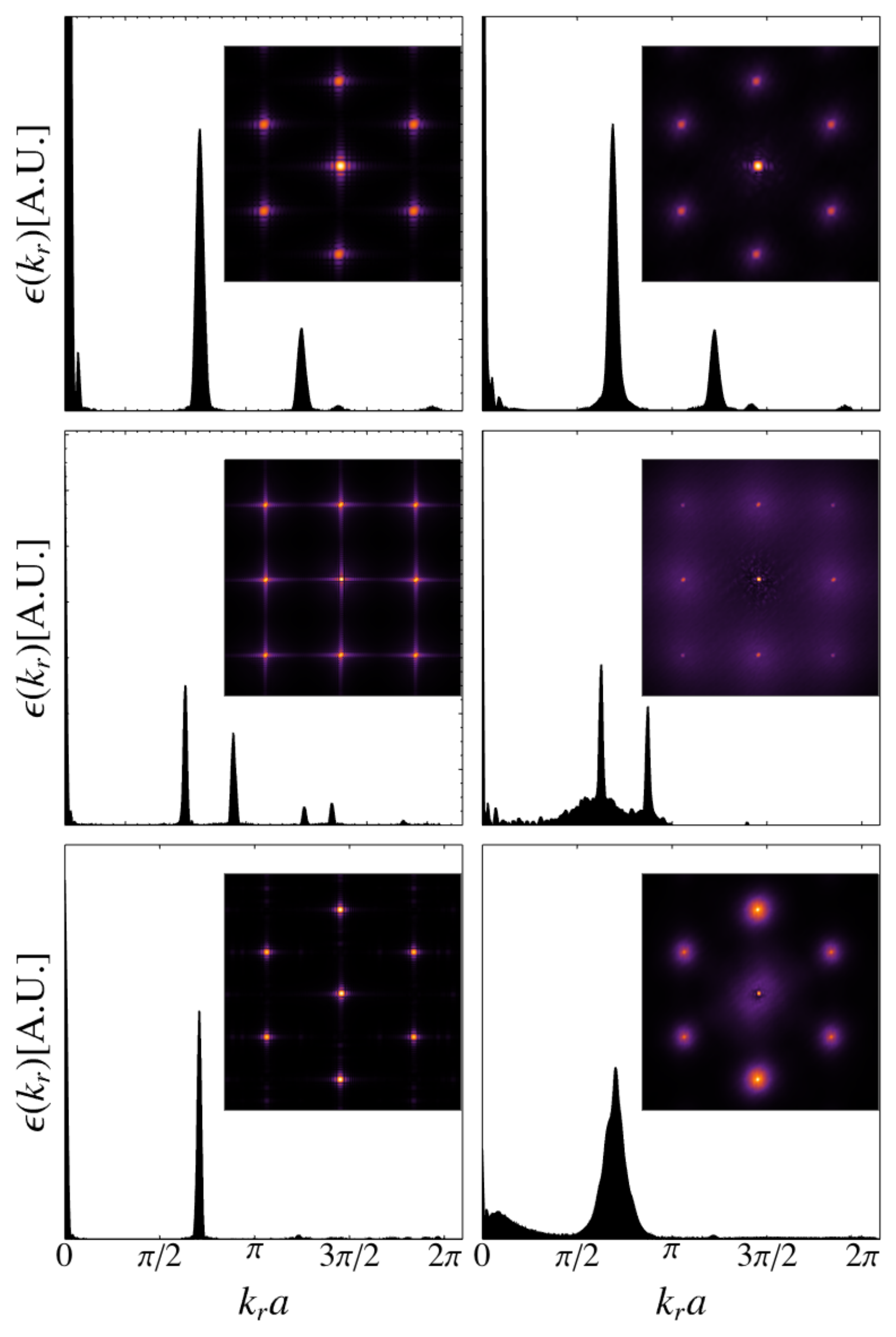}  
\end{center}
\caption{Energy spectrum in reciprocal space for honeycomb (top row), square (middle row) and triangular (bottom row) lattices in arbitrary units. Left and right columns correspond to disorder amplitudes of $\delta=0$ and $\delta = 80\%$ of the potential depth $V_0=12E_R$ respectively. Each of the energy spectra on the right column is the average over $\sim50$ realizations for a given disorder amplitude. The inset in each panel shows the corresponding density plot of the energy spectrum in the Brillouin zone.}
\label{Fig5}
\end{figure}

Analogously to the previous analysis for characterisation of density profiles, the ground state spectra analysis was done considering a significant number of realisations for a given disorder amplitude. Specifically, the results presented here correspond to the average over $\sim50$ realisations. For all of our numerical calculations, we use $\Omega\sim10^3$ and work within the first Brillouin zone (FBZ). For square lattices, this region is bounded by $-\frac{\pi}{a} \le k_x,k_y \le \frac{\pi}{a}$, while for honeycomb and triangular lattices the FBZ is the area of the primitive cell in the reciprocal lattice space generated by vectors ${\bf{b}}_1=\frac{4\pi}{a{\sqrt 3}}\left( \frac{\sqrt 3}{2}, -\frac{1}{2} \right)$ and ${\bf{b}}_2=\frac{4\pi}{a{\sqrt 3}}(0,1)$.  In Fig. \ref{Fig5}, we show the energy spectra of the lattices considered in our study for $\delta =0,80\%$ (left and right columns respectively) of the potential depth $V_{\mathrm{Latt}}^{\delta=0}(x,y)$, 
in the first quadrant of the FBZ. From top to down, we plot $\epsilon(k_r)$ vs. $k_{r}=\sqrt{k_x^2 +k_y^2}$ for honeycomb, square and triangular lattices respectively. The inset of each panel contains a density plot of ground state spectra in the full FBZ, associated to both, $\delta=0$ (left column) and $\delta = 80\%$  (right column), we should emphasize that left column corresponds to the averaged ground state spectra.  We observe that the system with disorder presents a ground state spectra which is no longer composed of sharp energy peaks, originated from the lattice symmetry. Disorder broadens the peaks in the distribution. In other words, instead of the sharp ground state spectra at zero disorder, we have a dense continuous behaviour when the disorder amplitude is different from zero. Interestingly, even though disorder is spanned across the entire lattice, and that this corresponds to the average over many realisations, we do not observe any cancellation effect due to randomness. Regarding the influence of the disorder strength, we found that as the disorder amplitude is increased the magnitude of the background increases until it becomes comparable to the amplitude of the signal due to the lattice symmetry. At this point, it is when the disorder effectively destroys the lattice symmetry, rendering the geometry effects indistinguishable. Summarising, the influence of disorder on the ground state spectra is manifested in replacing well defined peaks as a function of momenta with a super-imposed dense distribution due to incommensurability of spatial frequencies. 

\section*{Summary of findings and outlook}
We have studied the stationary states of ultracold weakly interacting bosonic atoms confined in two dimensional disordered optical lattices having different lattice geometries. In particular, we investigated the influence of energetic disorder in lattices with honeycomb (graphene like), square and triangular geometries, in producing localised states. The characterisation of localised states was performed by means of a systematic analysis through numerical simulations at mean-field level, using the Gross-Pitaevskii equation. The quantities analysed in our study, as a function of the disorder amplitude, were the average of the heights amplitudes and the fraction of peaks obtained directly from the density profiles, and the ground state spectra determined by Fourier transforming the energy associated to the ground state. For our numerical analysis, we considered lattices of size $\Omega \sim 10^3$ sites, constant effective on-site interaction strength $U=0.01 E_R$, and variable disorder amplitude in the interval $ \delta \in [0,1] E_R$, being these two quantities such that the Bose gas is well inside in the superfluid region. We also performed a finite size analysis considering the effects of the system size in the range $5 \times10^2 \lesssim \Omega \lesssim 5 \times10^3$. 

We found that localisation in two dimensional lattices occurs as a smooth crossover for square and triangular lattices. On the contrary, honeycomb lattices do not exhibit such localisation for the same disorder amplitudes and the same lattice amplitude $V_0/E_R$. This behaviour can be attributed to the characteristic larger tunneling energy of the honeycomb lattice with respect to that for square and triangular geometries. From the analysis of large lattice sizes, we found that starting from a certain disorder amplitude, the system exhibits partial localisation manifested as disconnected islands of density. We arrived at this conclusion from the analysis of the density since it saturates to a constant value as a function of the disorder strength. The original prediction made by Anderson is not completely restored for our weakly interacting disordered lattices in the sense that arbitrary disorder values do not produce localisation effects. Being this result analogous to that occurring in 1D, in which a weakly interacting Bose gas populates a finite number of localised single-particle Lifshitz states\cite{BGfrag}. However, localisation in few sites is found for large disorder strengths. Therefore, although localisation or density fragmentation occurs as a smooth crossover, we can identify regions with marginal localisation for $\delta\lesssim0.3$, weak localisation for $0.3\lesssim\delta\lesssim 0.6$ and strong localisation for $\delta\gtrsim 0.6$. We recognised such strong localisation as the collapse of the condensate density into a single density peak randomly located across the lattice, the limit when $p_f\to0$. Thus, transport is strongly suppressed for any lattice geometry for sufficiently large disorder strength. Our analysis also allowed to conclude that the coordination number determines how fast the smooth transition occurs as the system size increases, being the triangular lattice the first in which transport in inhibited against square and honeycomb lattices. It is interesting to note that although the honeycomb lattice do not constitute a bonafide Bravais lattice in 2D since it is composed by the superposition of two sub-lattices, our results are in good agreement with the fact that this lattice is an intrinsic good conductor.
 
From the analysis of the ground state spectra, we conclude that the influence of disorder is to replace a sharply peaked distribution for a continuous one. The disordered system ground state spectra is composed by continuous energy levels originated from disorder and the sharp energy level contribution originated from the underlaying symmetry of the lattice. In essence, the spatial random inhomogeneous distribution translates into broadening of the characteristic peaks of perfect lattices in the ground state energy distribution in momentum space.

Summarising, in our work we have been able to quantify the crossover to localisation as a function of the disorder amplitude for three different geometries. A general statement arising from our investigation is that disordered honeycomb lattices are robust against disorder as compared to triangular and square geometries for the same parameter ranges. As mentioned above, associated to the larger tunneling energy of the honeycomb lattices against their respective in square and triangular lattices, the honeycomb structure would have better conductivity in the presence of disorder in 2D. 

The present work provides some insight to understand localisation phenomena induced by disorder in two dimensional weakly interacting systems for lattices of different geometries having coordination $z=3,4$ and 6. It is interesting to note although Gross-Pitaevskii equation has an intrinsic mean-field nature, the quantities analysed to investigate localisation phenomena, namely the density distribution across the lattice and the ground state spectra, indeed exhibit signatures associated to the lattice geometry. Additionally, the information displayed in the ground state spectra in the reciprocal space can be compared with direct experimental measurements in ultracold systems via time of flight measurements~\cite{TOF}. Further investigations of transport phenomena can be addressed with the approach followed in the present work. For instance, it is possible to consider in a square lattice, different disorder amplitudes in  $\hat e_x$ and $\hat e_y$, that is, setting $\delta_x \neq \delta_y$. In view of the results obtained here, this could trigger the formation of stripes where transport could be manipulated via disorder induced localisation. Moreover, dynamical effects in adiabatic and non-adiabatic transfer protocols for the engineering of quantum states could be analysed\cite{LZ,LZ2}.  Beyond ultracold atoms, systems of ions\cite{Ions}, superconducting devices (Circuit-QED)\cite{CQED}, polariton systems\cite{PolHon,PolHon2} and the addition of high-Q cavities\cite{QOL} offer opportunities to test our findings analogously via quantum simulation\cite{QSim}. Moreover, the interplay of these effects in lattices with dissipation offer another venue of exploration\cite{Disipation, Disipation2} or measurement induced dynamics\cite{PRACab}. The findings of our work can aid in the development and characterise of transport properties in two dimensional systems for the development of new analog quantum technologies\cite{QTech}.

\section*{Acknowledgements}

This work was partially funded by grants IN105217, IN111516, IN109619 from DGAPA-UNAM, 255573 and A1-S-30934 from CONACYT. LAGG acknowledges scholarship from CONACYT. SFCB acknowledges support from C\'atedras CONACYT para J\'ovenes Investigadores project No. 551 and EPSRC Project EP/I004394/1. SFCB also thanks Consorcio CICESE-INAOE-CIO in PIIT, Apodaca N.L.,  IF-UNAM,  and the University of Oxford for their hospitality.

\section*{Author contributions statement}
LAGG performed the numerical experiments. SFCB supervised the numerical experiments and RP supervised the project.  All authors analysed and worked in the interpretation of the results.  All authors reviewed and wrote the manuscript. 

\section*{Additional Information}
\textbf{Competing financial interests.} The Authors declare no competing financial interests. \textbf{Competing non-financial interests.} The Authors declare no competing non-financial interests.

\newpage

\begin{thebibliography}{0}
\bibitem{RefU2D} This coupling constant is related to the standard 3D coupling interaction $g_{\mathrm{3D}}$, as $g_{\mathrm{2D}} \approx \frac{g_{\mathrm{3D}}}{l_z}$, with $g_{\mathrm{3D}}$ being the usual coupling interaction written in terms of the $s-$wave scattering length $a_{sc}$, $g_{\mathrm{3D}}=\frac{4 \pi \hbar^2 a_{sc} N}{m}$, and $l_z$ being the natural length of the ground state harmonic oscillator of a condensate originally confined in 3D and then squeezed along $z$ direction, thus resulting into a disk shape condensate confined in the $x-y$ plane.
\bibitem{fn2} {The size of the grid was chosen to ensure the norm conservation. The Split-Step Crank-Nicholson method in imaginary time was used to find the stationary state. To ensure the reliability of the stationary state found, we evolved such a state in real time and verified its invariance.}
\end{thebibliography}
\end{document}